
\input epsf                                                               %
\input harvmac
\def\Title#1#2{\rightline{#1}\ifx\answ\bigans\nopagenumbers\pageno0\vskip1in
\else\pageno1\vskip.8in\fi \centerline{\titlefont #2}\vskip .5in}

%
%
\ifx\epsfbox\UnDeFiNeD\message{(NO epsf.tex, FIGURES WILL BE IGNORED)}
\def\figin#1{\vskip2in}
\else\message{(FIGURES WILL BE INCLUDED)}\def\figin#1{#1}
\fi
\def\Fig#1{Fig.~\the\figno\xdef#1{Fig.~\the\figno}\global\advance\figno
 by1}
%
%
%
%
\def\ifig#1#2#3#4{
\goodbreak\midinsert
\figin{\centerline{\epsfysize=#4truein\epsfbox{#3}}}
\narrower\narrower\noindent{\footnotefont
{\bf #1:}  #2\par}
\endinsert
}
%
%
\def\rtg{\sqrt{g}}
\def\dfx{d^4x}
\def\fmn{F_{\mu\nu}}
\def\fMN{F^{\mu\nu}}
\def\calL{{\cal L}}
\def\osp{{1\over16\pi}}
\def\Rmn{R_{\mu\nu}}
\def\qh{{\hat q}}
\def\Bh{{\hat B}}
\def\Mpl{ {M_{\rm Planck} }}
\def\mgut{M_{\rm GUT}}

\def\RN{Reissner-Nordstr\o m}
\def\calo{{\cal O}}

\def\hf{{1\over2}}
\font\ticp=cmcsc10
\def\sq{{\vbox {\hrule height 0.6pt\hbox{\vrule width 0.6pt\hskip 3pt
   \vbox{\vskip 6pt}\hskip 3pt \vrule width 0.6pt}\hrule height 0.6pt}}}
\def\ajou#1&#2(#3){\ \sl#1\bf#2\rm(19#3)}
\def\frac#1#2{{#1 \over #2}}
\def\ie{{\it i.e.}}

\def\p+{{\partial_+}}

\def\ltwid{{\mathrel{\raise.3ex\hbox{$<$\kern-.75em\lower1ex\hbox{$\sim$}}}}}
\def\gtwid{{\mathrel{\raise.3ex\hbox{$>$\kern-.75em\lower1ex\hbox{$\sim$}}}}}
\def\Ssl{{\,\raise.15ex\hbox{/}\mkern-10.5mu S}}
\def\sq{{\vbox {\hrule height 0.6pt\hbox{\vrule width 0.6pt\hskip 3pt
   \vbox{\vskip 6pt}\hskip 3pt \vrule width 0.6pt}\hrule height
0.6pt}}}

\def\oep{{1\over8\pi}}
%
%
\lref\Erns{F. J. Ernst, \ajou J. Math. Phys. &17 (76) 515.}
\lref\Melv{M. A. Melvin, \ajou Phys. Lett. &8 (64) 65.}
\lref\GiPe{G.W. Gibbons and M.J. Perry, ``The physics of 2-d stringy
spacetimes,'' hep-th/9204090.}
\lref\Gibb{G.W. Gibbons, ``Quantized flux tubes in Einstein-Maxwell theory
and  noncompact internal
spaces,'' in {\sl Fields and geometry}, proceedings of
22nd Karpacz Winter School of Theoretical Physics: Fields and
Geometry, Karpacz, Poland, Feb 17 - Mar 1, 1986, ed. A. Jadczyk (World
Scientific, 1986).}
\lref\AfMa{I.K. Affleck and N.S. Manton, ``Monopole pair production in a
magnetic field,''\ajou Nucl. Phys. &B194 (82) 38.}
\lref\AAM{I.K. Affleck, O. Alvarez, and N.S. Manton,
``Pair production at strong
coupling in weak external fields,''\ajou Nucl. Phys. &B197 (82) 509.}
\lref\LNW{ K. Lee, V.P. Nair, and E.J. Weinberg, `` A classical
instability of Reissner-Nordstrom solutions and the fate of
magnetically charged black holes,'' Phys. Rev. Lett. {\bf 68} (92) 1100,
hep-th/9111045\semi
``Black holes in magnetic monopoles,'' Phys. Rev. {\bf D45} (1992) 2751,
hep-th/9112008.}
\lref\Japan{S.B. Giddings, ``Black holes and quantum predictability,''
UCSB preprint
UCSBTH-93-16, hep-th/9306041, to appear in the proceedings of the 7th
Nishinomiya Yukawa Memorial Symposium, K. Kikkawa and M. Ninomiya, eds.}
\lref\GiHa{G.W. Gibbons and S.W. Hawking, ``Action integrals and partition
functions in quantum gravity,''\ajou Phys. Rev. &D15 (77) 2752.}
\lref\Sredii{M. Srednicki, ``Entropy and area,'' LBL preprint LBL-33754.}
\lref\QTDG{S.B. Giddings and A. Strominger, ``Quantum theories of dilaton
gravity,''\ajou Phys. Rev. &D47 (93) 2454, hep-th/9207034.}
\lref\CBHR{S.B. Giddings, ``Constraints on black hole remnants,'' UCSB
preprint UCSBTH-93-08, hep-th/9304027.}
\lref\GiNe{S.B. Giddings and W.M. Nelson, ``Quantum emission from
two-dimensional black holes,''\ajou Phys. Rev. &D46 (92) 2486,
hep-th/9204072.}
\lref\DXBH{S.B. Giddings and A. Strominger, ``Dynamics of Extremal Black
Holes,''\ajou Phys. Rev. &D46 (92) 627, hep-th/9202004.}
\lref\CGHS{C.G. Callan, S.B. Giddings, J.A. Harvey, and A. Strominger,
``Evanescent black holes,"\ajou Phys. Rev. &D45 (92) R1005, hep-th/9111056.}
\lref\BGHS{B. Birnir, S.B. Giddings, J.A. Harvey, and A. Strominger,
``Quantum black holes,''\ajou Phys. Rev. &D46 (92) 638,
hep-th/9203042.}
\lref\Hawk{S.W. Hawking, ``Particle creation by black
holes,"\ajou Comm. Math. Phys. &43 (75) 199.}
\lref\Hawkii{S.W. Hawking, ``The unpredictability of quantum
gravity,''\ajou Comm. Math. Phys &87 (82) 395.}
\lref\BPS{T. Banks, M.E. Peskin, and L. Susskind, ``Difficulties for the
evolution of pure states into mixed states,''\ajou Nucl. Phys. &B244 (84)
125.}
\lref\Sred{M. Srednicki, ``Is purity eternal?,'' UCSB preprint
UCSBTH-92-22, hep-th/9206056.}
\lref\Cole{S. Coleman, ``Black holes as red herrings: Topological
fluctuations and the loss of quantum coherence,''\ajou Nucl. Phys. &B307
(88) 867.}
\lref\LoI{S.B. Giddings and A. Strominger, ``Loss of incoherence and
determination of coupling constants in quantum gravity,''\ajou Nucl. Phys.
&B307 (88) 854.}
\lref\CaWi{R.D. Carlitz and R.S. Willey, ``Reflections on moving
mirrors,''\ajou Phys. Rev. &D36 (87) 2327; ``Lifetime of a black
hole,''\ajou Phys. Rev. &D36 (87) 2336.}
\lref\Pres{J. Preskill, ``Do black holes destroy information?'' Caltech
preprint CALT-68-1819, hep-th/9209058.}
\lref\ACN{Y. Aharonov, A. Casher, and S. Nussinov, ``The unitarity
puzzle and Planck mass stable particles,"\ajou Phys. Lett. &B191 (87)
51.}
\lref\BDDO{T. Banks, A. Dabholkar, M.R. Douglas, and M O'Loughlin, ``Are
horned particles the climax of Hawking evaporation?'' \ajou Phys. Rev.
&D45 (92) 3607.}
\lref\BOS{T. Banks, M. O'Loughlin, and A. Strominger, ``Black hole remnants
and the information puzzle,'' hep-th/9211030, {\sl Phys. Rev. D} to
appear.}
\lref\GiMa{G.W. Gibbons and K. Maeda, ``Black holes and membranes in
higher-dimensional theories with dilaton fields,''\ajou Nucl. Phys. &B298
(88) 741.}
\lref\GaSt{D. Garfinkle and A. Strominger, ``Semiclassical Wheeler wormhole
production,''\ajou Phys. Lett. &256B (91) 146.}
\lref\GHS{D. Garfinkle, G. Horowitz, and A. Strominger, ``Charged black holes
in string theory,''\ajou Phys. Rev. &D43 (91) 3140, erratum\ajou Phys. Rev.
& D45 (92) 3888.}
\lref\MSW{G. Mandal, A Sengupta, and S. Wadia, ``Classical solutions of
two-dimensional
string theory,'' \ajou Mod. Phys. Lett. &A6 (91) 1685.}
\lref\Witt{E. Witten, ``On string theory and black holes,''\ajou Phys. Rev.
&D44 (91) 314.}
\lref\RST{J.G. Russo, L. Susskind, and L. Thorlacius, ``Black hole
evaporation in 1+1 dimensions,''\ajou Phys. Lett. &B292 (92) 13,
hep-th/9201074.}
\lref\Suth{L. Susskind and L. Thorlacius, ``Hawking radiation and
back-reaction,''\ajou Nucl. Phys & B382 (92) 123, hep-th/9203054.}
\lref\Hawkiii{S.W. Hawking, ``Evaporation of two dimensional black
holes,''\ajou Phys. Rev. Lett. & 69 (92) 406,
hep-th/9203052.}
\lref\RSTii{J.G. Russo, L. Susskind, and L. Thorlacius, ``The endpoint of
Hawking radiation,''\ajou Phys. Rev. &D46 (92) 3444, hep-th/9206070}
\lref\deAl{S.P. deAlwis, ``Quantization of a theory of 2d dilaton
gravity,''\ajou Phys. Lett. &B289 (92) 278, hep-th/9205069; ``Black
hole
physics from Liouville theory,''
Boulder preprint COLO-HEP-284, hep-th/9206020.}
\lref\Laws{For a review see R. M. Wald, 1991 Erice Lectures (unpublished). }
\lref\Bek{J. D. Bekenstein, ``Black holes and entropy,'' \ajou Phys. Rev. &D7
 (73) 2333 \semi ``Generalized second law of thermodynamics in black-hole
 physics,'' \ajou Phys. Rev. &D9 (74) 3292.}
\lref\PiSt{T. Piran and A. Strominger, ``Numerical analysis of black hole
evaporation,'' ITP preprint NSF-ITP-93-36, hep-th/9304148}
\lref\RuTs{J.G. Russo and A.A. Tseytlin, ``Scalar-tensor quantum gravity
in two dimensions,'' Stanford/Cambridge preprint
SU-ITP-92-2=DAMTP-1-1992.}
\lref\BiCa{A. Bilal and C. Callan, ``Liouville models of black hole
evaporation,'' Princeton preprint PUPT-1320, hep-th/9205089.}
\lref\tHoo{G. 't Hooft, ``The black hole interpretation of string
theory,''\ajou Nucl. Phys. &B335 (90) 138.}
\lref\Page{D. Page, ``Black hole information,'' Alberta-Thy-23-93, to
appear in the proceedings of the
5th Canadian Conference on General Relativity and Relativistic Astrophysics,
Waterloo, Ontario, 1993 May 13-15, hep-th 9305040.}
\lref\VeVe{E. Verlinde and H. Verlinde, ``A unitary S matrix and 2-d black
hole formation and evaporation,'' Princeton preprint PUPT-1380, hep-th
9302022.}
\lref\SVV{K. Schoutens, E. Verlinde and H. Verlinde, ``Quantum black hole
evaporation,'' Princeton preprint PUPT-1395.}
\lref\StTr{A. Strominger and S. Trivedi, ``Information consumption by
Reissner-Nordstrom black holes,'' ITP/Caltech preprint
NSF-ITP-93-15=CALT-68-1851, hep-th/9302080.}
\lref\RuTs{J.G. Russo and A.A. Tseytlin, ``Scalar-tensor quantum gravity
in two dimensions,''\ajou Nucl. Phys. & B382 (92) 259.}
\lref\HVer{H. Verlinde, lectures, this volume.}
\lref\PaSt{Y. Park and A. Strominger, ``Supersymmetry and positive energy in
classical and quantum
two-dimensional dilaton gravity,''\ajou Phys. Rev. &D47 (93) 1569.}
\lref\NePa{W. Nelson and Y. Park, ``N=2 supersymmetry in two-dimensional
dilaton gravity,'' UCSB preprint UCSBTH-93-10, hep-th/9304163.}
\lref\Noji{S. Nojiri, ``Dilatonic supergravity in two-dimensions and the
disappearance of quantum
black hole,''\ajou Mod. Phys. Lett. &A8 (93) 53, hep-th/9209118.}
\lref\MaSh{E. Martinec and S.L. Shatashvili, ``Black hole physics and
Liouville theory,''\ajou Nucl. Phys. &B368 (92) 338.}
\lref\Das{S.R. Das, ``Matrix models and black holes,''\ajou Mod. Phys.
Lett. &A8 (93) 69, hep-th/9210107.}
\lref\DMW{A. Dhar, G. Mandal, and S.R. Wadia, ``Stringy quantum effects in
two-dimensional black hole,''\ajou Mod. Phys. Lett. &A7 (92) 3703;
``Wave propagation in stringy black hole,'' Tata preprint TIFR-TH-93-05,
hep-th/9304072.}
\lref\JeYo{A. Jevicki and T. Yoneya, ``A deformed matrix model and
the black hole background in two-dimensional string
theory,'' ITP preprint NSF-ITP-93-67=Komaba preprint
UT-KOMABA 93/10, hep-th/9305109.}
\lref\HaSt{J.A. Harvey and A. Strominger, ``Quantum aspects of black
holes,'' preprint EFI-92-41, hep-th/9209055, to appear in the proceedings
of the 1992 TASI Summer School in Boulder, Colorado.}
\lref\Erice{S.B. Giddings, ``Toy models for black hole evaporation,''
UCSBTH-92-36, hep-th/9209113, to appear in the proceedings of the
International Workshop of Theoretical Physics, 6th Session, June 1992,
Erice, Italy.}
\lref\Dyso{F. Dyson, Institute for Advanced Study preprint, 1976,
unpublished.}
\lref\Astro{A. Strominger, ``Fadeev-Popov ghosts and 1+1 dimensional black
hole evaporation,'' UCSB preprint UCSBTH-92-18, hep-th/9205028.}

\Title{\vbox{\baselineskip12pt\hbox{UCSBTH-93-17}
\hbox{gr-qc/9306023}
}}
{\vbox{\centerline { Entropy in}\vskip2pt
\centerline{Black Hole Pair Production}
}}

\centerline{{\ticp David Garfinkle}${}^{1,3}$}
\centerline{{\ticp Steven B. Giddings}${}^{2,1}$\footnote{$^\dagger$}{Email
addresses:
giddings@denali.physics.ucsb.edu, steve@voodoo.bitnet.} }
\centerline{{\ticp Andrew Strominger}${}^{1,2}$ }
\vskip.1in

\centerline{\sl ${}^1$Institute for Theoretical Physics}
\centerline{\sl and}
\centerline{\sl ${}^2$Department of Physics}
\centerline{\sl University of California}
\centerline{\sl Santa Barbara, CA 93106}
\vskip.1in
\centerline{\sl ${}^3$Department of Physics}
\centerline{\sl Oakland University }
\centerline{\sl Rochester, MI 48309 }
\bigskip
\centerline{\bf Abstract}
{Pair production of \RN\ black holes in a magnetic field
can be described by a euclidean instanton. It is shown that
the instanton amplitude contains an explicit factor
of $e^{A/4}$, where $A$ is the area of the event horizon.
This is consistent with the hypothesis that  $e^{A/4}$
measures the number of black hole states.}
\Date{}

\newsec{Introduction}

The elegant laws of black hole thermodynamics\refs{\Bek,\Laws}
have yet to find a
microscopic
explanation in an underlying statistical mechanics
of black hole states.  Particularly interesting is the
interpretation of the Bekenstein-Hawking  entropy.
In most cases this entropy is given by the simple formula
\eqn\alaw{S_{\rm bh}=A/4,}
where $A$ is the area of the black hole horizon in Planck units. Up to an
additive constant, this formula can be derived by
insertion of the semiclassical black hole mass-temperature
relation\refs{\Hawk} into the thermodynamic
formula $1/T=\partial S / \partial E$
followed by integration. Additional assumptions are
required to fix the constant
part of the entropy. A widely utilized,  but mysterious, procedure is to fix
the constant by relating it to the black hole instanton in the Euclidean
path integral\refs{\GiHa}.

The relation \alaw\ acquires additional meaning in light of Bekenstein's
conjectured generalized second law\refs{\Bek}, which states that the sum of the
usual entropy plus $S_{\rm bh}$ always increases. Although there
is no complete proof of this conjecture,
evidence is provided by the many ingenious
gedanken attempts\refs{\Laws} to violate the generalized second law
which have been foiled by the subtle dynamics of quantum mechanical black
holes.

If the
traditional connection between thermodynamics and statistical mechanics
were to extend to black holes, then the number of
quantum states of the black hole would be finite and given
by
\eqn\bhstates{N=e^{S_{\rm bh}}\ .}
These microstates might be either ``internal states'' inside the black
hole or ``horizon states'' somehow associated with degrees of freedom
of (or near) the black hole horizon, or both.

The issue of whether \bhstates\
can be taken literally has bearing
on the vexing question of what happens to information cast into a
black hole\foot{For recent reviews see \refs{\Pres\HaSt\Erice-\Japan}.}.
If one assumes that \bhstates\ counts {\it all} the black hole states,
and that information is preserved, then one is forced to conclude that
information escapes from a black hole at a rapid rate (proportional to
the rate of area decrease) during the
Hawking process. We do not think this is likely because it seems to
requires a breakdown of semiclassical methods for arbitrarily large
black holes and at arbitrarily weak curvatures, although this
point is certainly the subject of heated debates! On the other hand one might
try to account for the decrease in \bhstates\ during black hole evaporation
by assuming that information is truly lost in the black hole interior,
perhaps being eaten by the singularity. A problem with this is that -
for large neutral black holes - the spacelike slice on which the
quantum Hilbert space is defined can be extended through the interior
of the black hole in a manner which avoids the singularity and all
strong-curvature regions. Dynamics on such a slice is weakly
coupled, and it is therefore hard to see how information could be
lost\foot{Although
perhaps \bhstates\ makes sense only with respect to a specific slicing of
spacetime which differs from the one described here.}.

An alternate interpretation of \bhstates\ is that it counts only
the horizon states. One is then not pushed into the conclusion
that information either rapidly escapes or is eaten at weak coupling.
Indeed if one assumes that there are $e^{1/4}$
states per Planck area of the horizon, one precisely recovers \bhstates.
Certainly a derivation of this strange factor would be of great interest!
Of course even if such a derivation were found, it would still remain to
understand why - if \bhstates\ counts only horizon states - the
generalized second law appears to be valid.

Yet a third microphysical explanation of \alaw is suggested by
recent work\refs{\Sredii}, in which the entropy of the free scalar
field vacuum
outside a ball of surface area $A$ was computed by tracing over
states inside the ball. The result was found to leading order in $A$
to be
\eqn\srent{S_{\rm bh}=A \Lambda^2,}
where $\Lambda$ is an ultraviolet cutoff. \srent\ has a microphysical
explanation by construction, but it is not in terms of states at
or inside the surface of the ball. Rather the entropy
arises from correlations between the quantum state inside and outside
the ball. It is tempting to try to relate this observation to
\alaw, but this would require explaining why $ \Lambda^2$ is
precisely $1/4$ in Planck units. Furthermore,
such an interpretation of \alaw\ would not appear to
readily explain the validity of the generalized second law.
Certainly no such law is valid in the free field example of \refs{\Sredii}.


For these reasons it is clearly of interest to seek a deeper understanding
of the meaning of the black hole entropy.  One promising avenue of
exploration is the phenomenon of pair production of charged black holes.
In Schwinger production of charged particles in a
background field, the total production rate grows as the number of particle
species produced.  If this is extrapolated to black hole production
in a background field\refs{\Gibb,\GaSt} then one would likewise expect the
rate to be proportional to the number of independent black hole
states produced. In this paper we show that the factor \bhstates\
indeed multiplies the pair production amplitude, consistent with
its interpretation as somehow counting black hole microstates.
While the nature of these supposed states is still very mysterious,
we do hope that our result will constrain future interpretations.

The desired factor \bhstates\ is isolated from the rest of the
pair production amplitude by consideration of the family of stable solutions
discussed in \refs\LNW\ corresponding to gravitationally corrected
`t Hooft-Polyakov monopoles of charge $q$. For $qM_{GUT} \ll \Mpl$, these
closely resemble the `t Hooft-Polyakov solutions.  For $qM_{GUT}> \Mpl$,
the monopole drops inside an event horizon and the
solutions are identical to extremal Reissner-Nordstrom monopole black holes.
Pair production of these monopoles can be analyzed using instanton
methods.  For fixed magnetic field $B$, consider a one-parameter family of
instantons labeled by $M_{GUT}$. For $qM_{GUT} \ll \Mpl$, the instanton
resembles
the one described by Affleck, Alvarez, and Manton \refs{\AfMa,\AAM} as an
`t Hooft-Polyakov monopole in a circular orbit in euclidean space.
For $qM_{GUT}> \Mpl$, the  instanton is precisely the
one found in \refs{\GaSt} describing  Reissner-Nordstrom monopole
pair production. At the critical value of $M_{GUT}$ near $\Mpl/q$,
where the monopole drops inside a horizon, one finds that the action
discontinuously changes by precisely $-S_{\rm bh}$.

Of course even our well-funded gedanken experimentalist can not
observe this threshold because coupling constants such as $M_{GUT}$
cannot be varied in the laboratory. Fortunately it will be seen from a precise
description
of the production process that the same threshold
can be observed by varying the magnetic field
$B$ while keeping $M_{GUT}$ fixed. Our gedanken experimentalist who
discovers that the production rate suddenly jumps up at precisely
the critical $B$ field which produces monopoles with horizons,
will likely conclude that he has crossed a threshold for production of
$e^{S_{\rm bh}}$ new states. This assigns a new, physical
significance to the relation
\bhstates.

In section two we briefly review ref.~\refs{\GaSt}
and present an exact formula for the pair production rate.  It is however
somewhat difficult to extract from this the contribution of the entropy
because structure dependent Coulomb terms give contributions of similar
magnitude.  To circumvent this difficulty, section three compares
this amplitude to
the pair production of a GUT
monopole (with parameters tuned so that its surface is barely outside the
would-be
horizon) and thereby extracts the entropy factor.
Finally, in section four we
perform the same comparison in the two-dimensional reduced theory that
arises in the weak-field limit.  Although this yields exactly
the same result, it
provides a simplified description of the process.  Section five closes with
discussion. The appendix contains a derivation of the exact action of the
black hole pair-production instanton, which is valid even for black holes
of size (or charge)
comparable to $1/B$. This extends the leading-order-in-$B$
expression given in \refs{\GaSt}.

\newsec{\RN\ pair production}

The amplitude for production of magnetically charged black holes  in a
magnetic field can be
calculated in the semiclassical approximation by finding an
analogue of the Schwinger instanton in gravity coupled to electromagnetism,
with euclidean action
\eqn\action{S=\osp\int d^4x \sqrt{g} \left[-R +
F_{\mu\nu}F^{\mu\nu}\right] - {1\over8\pi}\int d^3 x \sqrt{h} K\ }
where we have included the surface term written in terms of the extrinsic
curvature $K$ and boundary metric $h$.
First consider
the solution corresponding to the background field.  Because of the
magnetic energy this solution is not flat, but rather for a magnetic field
in the $z$ direction is given by the euclidean Melvin universe\refs{\Melv},
\eqn\Melvin{\eqalign{ds^2 &= (1+{1\over4} B^2 \rho^2)^2(dt^2 + dz^2
+d\rho^2) + {\rho^2\over(1+{1\over4} B^2 \rho^2)^2}d\phi^2\cr
F&= {B\rho d\rho\wedge d\phi\over (1+{1\over4} B^2 \rho^2)^2}}}
where $-\infty<t,z<\infty$, $0<\rho<\infty$, and $0<\phi<2\pi$.
This solution corresponds to a flux tube with total flux
\eqn\fluxe{\Phi=\int F = {4\pi\over B}}
through a transverse hypersurface.

The instanton describes circular motion
of an extremal \RN\ black hole in the euclidean continuation
of the Melvin universe.
It is given by the Ernst solution\refs{\Erns},
\eqn\Ern{\eqalign{
ds^2&= {\Lambda^2\over A^2 (x-y)^2} [-G(y)dt^2 - G^{-1}(y) dy^2 +
G^{-1}(x) dx^2] + {G(x)\over \Lambda^2 A^2 (x-y)^2 }dz^2\cr
F&=dz\wedge dE\ .}}
Here
\eqn\Gdef{G(x) = 1-x^2(1+\qh Ax)^2\ ,}
\eqn\Edef{E= {2\over\Lambda \Bh}(1+\hf \qh\Bh x)\ ,}
\eqn\Ldef{\Lambda = (1+ \hf \qh\Bh x)^2 + {\Bh^2 G(x) \over 4 A^2 (x-y)^2}\ ,}
and $A$, $\qh$ and $\Bh$ are parameters obeying
\eqn\constr{\qh\Bh<1/4\ .}
The function
$G$ then has four zeroes, $\zeta_1,\ldots,\zeta_4$;
$y$ is taken to run between $\zeta_2$ and $\zeta_3$ and $x$ runs between
$\zeta_3$ and $\zeta_4$.  As these zeroes are
approached the metric becomes singular unless periodic identifications are
made on $t$ and $z$.  This also forces a relation between $A$, $\qh$, and
$\Bh$,\foot{For more details, see the appendix.}
\eqn\abrelnx{1 = \left ( {{1 \; + \; 4 \qh A } \over {1 \; - 4 \qh A }}
\right ) \; {{\left (
{{ 1 \; + \; {1 \over 2} \qh \Bh {\zeta_3}  }
\over { 1 \; + \; {1 \over 2} \qh \Bh {\zeta_4} }} \right ) }^8} \ .}
This solution has topology $S^2\times S^2-\{p\}$ where $p$ corresponds to
$x=\zeta_3=y$.
The Melvin metric is recovered asymptotically as $x,y\rightarrow\zeta_3$.
A schematic picture of the solution is shown in fig.~1.

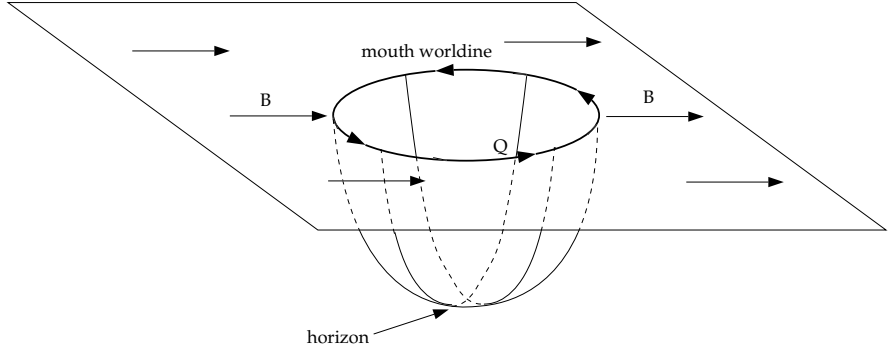
\ifig{\Fig\ernfig}{Shown is a schematic representation of the Ernst
solution.  The topology of the solution is that of $R^4$ asymptotically,
but internally has a circulating wormhole mouth.  Any given point in the
``cup'' region corresponds to a two sphere.}{ebhpp.fig1}{2.2}

By comparing the asymptotic form of the metric to the Melvin metric, the
value of the asymptotic magnetic field can be identified,
\eqn\magfield{B= \Bh \;  {{\left ( 1 \; + \; {1 \over 2} \;
\qh \Bh {\zeta_3} \right ) }^{ - 3}} \; {\sqrt { 1 \, - \, 4 \qh A}} \ .}
The value of the charge is found by computing the integral of $F$ over the
the two-sphere parametrized by $x$ and $z$ that encircles the horizon of the
black hole; this gives
\eqn\qcharge{\eqalign{q&= {1 \over {4 \pi}} \; {\int _{S^2}} \; { F}\cr
&={1 \over { B}} \; \left [ 1 \; - \; {{ 1 \; + \; {1 \over 2} \qh \Bh
{\zeta_3}  } \over { 1 \; + \; {1 \over 2} \qh \Bh {\zeta_4} }} \right ]\ .}}
For production of point particles of charge $q$ and mass $M$ the radius of
the circle on which the particles travel is given by
\eqn\mrad{l\sim {m\over qB}\ .}
In the case of extremal
black holes this radius should be bigger than that of the
black hole horizon, which is enforced by \constr.  The analogue of the
point particle limit is given by $qB\ll 1$.  We also want $q>1$ so that
the black holes are larger than the Planck radius.

To interpret the instanton's contribution to the production process it must
be cut in half along the moment of time symmetry surface
given by $t=$ constant.  The three geometry of this surface is that of a
wormhole, with a trapped magnetic flux.  The opposite ends of the wormhole
correspond to the pair of black holes.  Subsequent evolution arises from
continuation to lorentzian signature; the pair of mouths run away to
opposite ends of the magnetic field.

To evaluate the semiclassical production rate we need the instanton action.
The action is computed by calculating its change
\eqn\dinstact{{\delta S\over\delta q}}
under an infinitesimal variation of the charge of the black
hole.  This can then be integrated from zero to $q$ to give the total
action, as described in the appendix.  The result is
\eqn\instact{S = 4 \pi {{ q}^2} \; {{{\left ( 1 \, - \, { B} { q} \right )
}^2} \over {1 \; - \; {{\left ( 1 \, - \, { B} { q} \right ) }^4}}}\ .}
Expanding in the small parameter $qB$ we find
\eqn\expact{S= {\pi q\over B} - {\pi\over2} q^2 + \calo(q^3B)\ .}
The semiclassical production rate is thus
\eqn\rate{e^{-S} \sim e^{-{\pi q\over B}+ {\pi\over2} q^2+ \calo(q^3B)\ } }
up to factors arising from loops.  The leading term
corresponds precisely to the Schwinger rate, $\exp\{-\pi m^2/qE\}$ for pair
production in an electric field.


The subleading term in \rate\ is of the correct
order of magnitude to correspond to the black hole entropy, $S_{\rm bh} =
A/4 = \pi q^2$.  However, things are not so simple as various loop
corrections enter at the same order.  In particular there are Coulomb
corrections that arise from exchange of a photon line from one point on the
trajectory to another.  In the case of pair production of magnetic
monopoles, considered in \refs{\AfMa,\AAM}, these give a contribution
$\exp\{\pi q^2\}$.  In the present case one likewise expects a
gravitational Coulomb correction of the same size.  However, in contrast to
the monopole, in the present case there is not a well-defined
four-dimensional effective
field theory in which excitations around
the black hole can be ignored\foot{There is presumably an effective field
theory which contains two-dimensional regions corresponding to the long
throat outside the black hole. Such effective field theories were
derived for dilatonic black holes in \refs{\BDDO,\DXBH} and used to
analyze pair production in \refs{\BOS,\CBHR}.
It will be partially described for the
\RN\ case in section 5. }
Therefore there can be structure dependent corrections to these results.
This makes it difficult to compute the expected production rate and compare
it to the rate \rate\ to extract factors corresponding to the number of
states.

A second problem is that the approach of this section requires integrating
the action up from $q=0$.  One might worry that there could be a surface
term, perhaps even infinite, arising from the endpoint of the integration.
In the following section we will perform a different calculation that
addresses both of these problems.

\newsec{Comparison to monopole production}

In view of the various contributions to the semiclassical rate one needs a
better standard of comparison to extract the state-counting factor.  Such a
standard can be had by reconsidering the pair production of magnetic
monopoles in a grand unified theory with gravity included.  The mass and
size of
such a monopole are of order $q^2 \mgut$ and $1/\mgut$ respectively,
where $\mgut$ is the GUT mass
scale.  If we consider tuning the parameters so that $\mgut$ approaches
$\Mpl/q$, then the surface of the monopole barely hovers outside the
would-be
horizon.  Thus the solution is essentially \RN\ until very close to the
horizon.  Pair production of these objects is described by an instanton
like that of the preceding section, except the geometry near the bottom of
the ``cup'' of fig.~1 is cut off and replaced by the monopole circulating
around the loop, as shown in fig.~2.  All structure-dependent corrections
are therefore
the same in the two solutions and we can compute the structure
independent difference between black hole production and monopole
production rates.

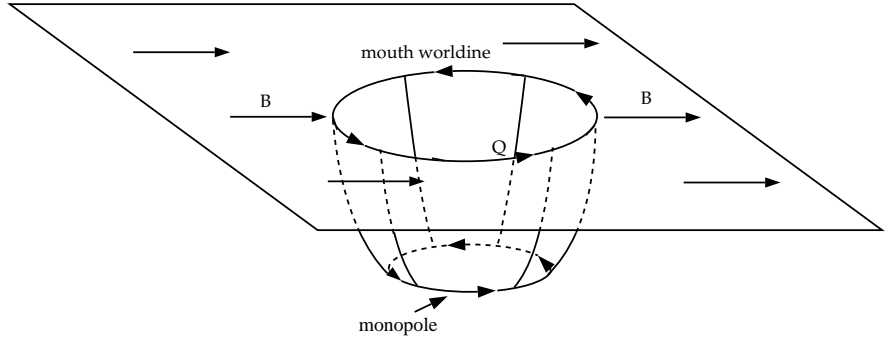
\ifig{\Fig\moncr}{Shown is the instanton of
fig.~1, but with the lower portion of the
cup truncated.  The resulting boundary corresponds to the $S^2$ surface of
the monopole, moving on a circular trajectory.  This instanton thus
describes pair production of gravitationally-corrected
't Hooft-Polyakov monopoles.}{ebhpp.fig2}{1.8}

As one increases the magnetic field, the radius of the euclidean orbit
decreases. Both the radius of the cup, and of the hole at the bottom of the
cup in fig. 2 will decrease. At a critical value of the $B$ field, the
hole closes up, because sufficient acceleration of a monopole
with $M_{GUT}$ just below $\Mpl$ results in a horizon.\foot{Note that
this provides a strong argument that topology change {\it must} be
included in quantum gravity. Otherwise there is no pair production instanton
above the critical value of $B$.}
Thus the difference of the actions above and below the critical value
can be interpreted as the threshold factor measured by a gedanken
experimentalist who varies $B$.

To proceed we must calculate the difference between the classical actions
for the respective processes.  The difference arises solely from the
difference between the actions of the monopole core and of the section of
the bottom of the cup that it replaces.  The latter tends to zero as the
core is tuned to the horizon, so all we need is the monopole action.

The latter is given by
\eqn\monact{S_m = \int d^4x\sqrt{g}\left[ -{R\over16\pi} + \calL_m\right]}
where the integral is over the monopole core and
$\calL_m$ is the lagrangian for gauge fields minimally coupled to
Higgs fields that break the GUT group to $U(1)$ (we do not need the
explicit form of this action).  The full solution has a Killing vector
$k^\mu$ that generates $t$ translations, \ie, rotations about the cup's
axis.  Using $k^\mu$ we will rewrite \monact\ as a surface
term.

Double contraction of Einstein's equations with this vector gives
\eqn\ecc{\oep k^\mu k^\nu \Rmn = k^\mu k^\nu T_{\mu\nu} + \osp k^2
R\ ,}
where $k^2=k^\mu k_\mu$.  The term involving the stress tensor is
\eqn\stress{k^\mu k^\nu T_{\mu\nu} = 2\left({\partial \calL_m\over\partial
g^{\mu\nu}}k^\mu k^\nu - \hf k^2 \calL_m\right) \ .}
In this expression the first term vanishes.  To see this, note that for the
static monopole configuration all time derivatives vanish, and furthermore
a gauge may be chosen so that only spatial components of the gauge field
are non-vanishing.  This means that the lagrangian cannot depend on
$g^{tt}$, as there are no indices for this to contract with.  Combining
\ecc\ and \stress\ therefore gives the lagrangian,
\eqn\euclag{ -{R\over16\pi} + \calL_m = -{1\over8\pi k^2} k^\mu k^\nu \Rmn
\ .}
The latter expression can be written as a total derivative:
the identity
\eqn\idone{ \nabla_\nu\nabla^\nu k_\mu =-k^\nu R_{\mu\nu} }
for Killing vectors gives
\eqn\rone{- k^\mu k^\nu \Rmn = k^\mu \nabla_\nu\nabla^\nu k_\mu\ .}
Then the hypersurface-orthogonality of $k^\mu$ implies
\eqn\idtwo{\nabla_\mu k_\nu = -k_{[\mu} \nabla_{\nu]} \ln k^2 }
which can be used to show
\eqn\rtwo{ {1\over k^2} k^\mu \nabla_\nu\nabla^\nu k_\mu= \hf \sq \ln k^2\
.}

The action \monact\ now becomes
\eqn\monsurf{S_m= \osp \int d\Sigma n^\mu \nabla_\mu \ln k^2}
where the integral is over a surface just outside the monopole core.  This
integral is readily evaluated using the approximate form
\eqn\approxm{ds^2 = r^2 d\tau^2 +dr^2  + h_{ij}dx^i dx^j\ }
of the metric
near the horizon; here $x^i$ parameterize the horizon two-sphere.
In these coordinates $k^\mu = (1,0,0,0)$ and we find
\eqn\sfin{S_m = {1\over4} \int d^2x \sqrt{h} = {1\over4}A\ }
where $A$ is again the horizon area.

Eq.~\sfin\ is precisely the usual black hole entropy.  Monopole
production is therefore suppressed relative to
black hole production, and the
ratio of the black hole production rate to that for monopoles is given by
\eqn\bhrat{{\Gamma_{\rm bh}\over\Gamma_{\rm m}} = e^{S_{\rm bh}}\ .}
This is exactly as one would expect if in creating black hole pairs one is
allowed to create an extra number of states given by \bhstates.

The surface term \sfin\ could equally well be understood by eliminating the
monopole core and including the Gibbons-Hawking\refs{\GiHa} surface term
in the action\foot{A related observation has been made by F. Wilczek
(private communication).}.  We have given the above derivation in order to make
the origin of this term more apparent.  In the next section we will
reproduce the same result from the latter approach, but will instead work
in the two-dimensional theory describing the geometry of the cup
portion of the instanton at small $B$.

\newsec{Production in the two-dimensional effective theory}

For weak magnetic field the above instanton and the contribution of the
black hole entropy can be approximately described in a two-dimensional
reduced theory.  To see this, recall that the spatial geometry of the
extremal \RN\ solution near the horizon is that of an infinitely long
throat. The
latter statement follows from making the redefinition
\eqn\redef{r-q= q e^{w}}
on the euclidean \RN\ metric
\eqn\rnmet{ds^2= \left(1-{q\over r}\right)^2 dt^2 +
{1\over\left(1-{q\over r}\right)^2} dr^2 + r^2 d\Omega_2^2\ .}
The limit $w\rightarrow -\infty$ corresponds to the vicinity of the horizon,
and in this limit the metric becomes
\eqn\rnlim{ds^2 = e^{2w}dt^2 + q^2 dw^2 + q^2 d\Omega_2^2\ .}
We will work in the approximation where all but the s-wave excitations are
dropped, and thus the angular directions will be ignored.
The corresponding two dimensional
euclidean action is found by dimensional reduction:
\eqn\seff{S_2= -{1\over4}\int d^2 \sigma \sqrt{g}\left[ e^{-2\phi}R+ 2
e^{-2\phi}\left(\nabla\phi\right)^2  +2-2q^2 e^{2\phi} \right]\ ,}
where
\eqn\metred{ds^2 = g_{\alpha\beta} d\sigma^\alpha d\sigma^\beta +
e^{-2\phi} d\Omega_2^2\ .}

A two-dimensional limit also occurs in the instanton \Ern.  In the limit
$B\rightarrow 0$ the radius of the black hole trajectory grows as $1/B$,
and the size of the cup in fig.~1 goes to infinity while the radius of the
horizon is fixed.  Therefore over the cup region the low-momentum theory is
effectively two-dimensional.
The vicinity of the
horizon is given by values of $y$ close to $\zeta_2$, and in this vicinity
the metric \Ern\  takes the form
\eqn\limmet{ ds^2= q^2\sinh^2w dt^2 + q^2 dw^2 + q^2 d\Omega_2^2 \ ,}
where we have defined $y-\zeta_2 = \cosh w-1$.  This is readily seen to
give a solution to \seff.

The effects of monopoles are incorporated in the two dimensional theory by
adding a boundary corresponding to the monopole worldline. The monopole
dynamics are encoded both in the boundary conditions and by the addition of
operators integrated along the boundary. All operators consistent
with the symmetries are expected to be present.

Monopole pair production as in fig.~2 corresponds
to terminating the surface at a circle just outside the horizon.  By tuning
the parameters  this boundary can be taken to be at any
value of $w$ near the horizon at $w=0$.  This can for example be
explicitly seen from the  boundary term
\eqn\surdens{ C \oint dl}
corresponding the the monopole energy.  Allowing variations
of the fields at the boundary will yield an equation fixing the
boundary radius in terms of $C$. In general $C$ and the coefficients
of the boundary operators are hard to compute.  Fortunately, there
is only one term which has a non-vanishing contribution as the
boundary circle shrinks to a point: namely
\refs{\GiHa,\GiPe}
\eqn\surft{ -{1\over2} \oint dl e^{-2\phi} k\ ,}
where $k$  is the extrinsic curvature of the surface.
Its coefficient must be precisely as in \surft, otherwise
the variational principle is not well-defined due
to surface terms involving the variation of the derivative of the metric at
the boundary.

In comparing the
action for monopole production to that for black hole production, the
surface term \surft\ contributes only in the former case.  Evaluating the
surface
term in that case gives $\pi q^2$, in agreement with \sfin.

\newsec{Discussion}

The above results show that pair production of \RN\ black holes is enhanced
by an extra factor of $e^{S_{\rm bh}}$ over that of pair production of
magnetic monopoles.  This is consistent with the black hole
entropy serving as a measure of the number of internal states.  However,
for several reasons it is far from clear why the black hole entropy is playing
this role.  Indeed, our calculations only include the classical action, and
not the functional determinant from the functional integral.  The latter is
expected to count states corresponding to fluctuations about the
semiclassical geometry (for
further discussion see \refs{\CBHR}).  We do not understand why
contributions from the {\it classical} action should
provide a factor appearing to count states, or how
these states might be described.  These matters deserve further
exploration.

\appendix{A}{Calculation of exact euclidean action}

In this appendix we will derive the result \instact\ for the exact action
of the instanton.  As outlined in the text, this is computed by first
finding the its variation under a small change in the charge of the black
hole, then integrating.  As stated
in \refs{\GaSt}, the variation of the gravitational part of the action
vanishes,
and we have
\eqn\vareqn{\eqalign{ \delta S &= {1\over8\pi}\int \dfx \rtg \fMN
\delta\fmn \cr
&= {1\over4\pi} \int \dfx\rtg\fMN \nabla_\mu \delta A_\nu\cr
&= {1\over4\pi} \int d^3x \sqrt{h} n_\mu \fMN \delta A_\nu
 }}
where the latter is a boundary integral with induced metric $h$.  This
boundary integral receives contributions only from the failure of $\delta
A$ to match on the equator of the horizon two sphere, and gives
\eqn\diffact{\delta S = I \delta { q}}
where
\eqn\Idef{I = {\int _{S^2}} \; * F \ .}
and $S^2$ is here the ``orbital'' two sphere.

First we calculate $I$.  From \Ern\ we find
\eqn\gone{{\int _{S^2}} \; * F = - \; {\int _{t_-} ^{t_+}} \; d t \; {\int
_{\zeta_2} ^{\zeta_3}} \; d y \; {\Lambda ^2} \; {{\partial E} \over {\partial
x}} \; \; \; .}
Note that this integral is evaluated on the equator of the horizon 2-sphere,
that is at
$ x = 0 $.
Note also that
$ G ( 0 ) = 1 $
and
$ G ' ( 0 ) = 0 $.
It then follows from the formulas \Edef, \Ldef\ for
$ E $
and
$ \Lambda $
that at
$ x = 0 $
we have
\eqn\gartwo{
{\Lambda ^2} \; {{\partial E} \over {\partial x}} = - \, \qh \; + \; {{\qh
{\Bh^2}}
\over {4 {A^2} {y^2}}} \; - \; {\Bh \over {{A^2} {y^3}}} \; \; \; .}
It then follows that
\eqn\gthree{\eqalign{
I =& \Delta t \; {\int _{\zeta_2} ^{\zeta_3}} \; d y \; \left ( \qh \; - \;
{{\qh {\Bh^2}}
\over {4 {A^2} {y^2}}} \; + \; {\Bh \over {{A^2} {y^3}}} \right )\cr
=& \Delta t \; \Delta y \; \left ( \qh \; - \; {{\qh {\Bh^2}}
\over {4 {A^2} {\zeta_3} {\zeta_2}}} \; + \; {{\Bh \, \left ( {\zeta_3} +
{\zeta_2} \right ) } \over { 2 {A^2} {{\left ( {\zeta_3} {\zeta_2} \right )
}^2} }} \right ) \; \; \; .
}}

The zeroes $\zeta_2$ and $\zeta_3$ are explicitly found to be
\eqn\zeroeS{\zeta_{2,3}= {1 \over {2 \qh A }} \; \left [ - 1 \; \mp \;
{\sqrt { 1 \, - \, 4 \qh A }} \right ] \; \; \; .}
It then follows that
\eqn\gfour{
I = \Delta t \; {\sqrt { 1 \, - \, 4 \qh A }} \; \left ( {1 \over A} \; - \;
{\Bh \over {2 {A^2}}} \; - \;  {{\qh {\Bh^2}} \over {4 {A^2}}} \right ) \;
\; \; .}
The quantity
$ \Delta t $
must have the value that makes the metric well behaved at the poles of the
orbital 2-sphere.  This gives
\eqn\gfive{
\Delta t = {{2 \pi } \over {\sqrt { 1 \, - \, 4 \qh A }} } \; \; \; .
}
So we find
\eqn\gsix{
I = 2 \pi \;  \left ( {1 \over A} \; - \;
{\Bh \over {2 {A^2}}} \; - \;  {{\qh {\Bh^2}} \over {4 {A^2}}} \right ) \;
\; \; .}

Unfortunately our expression for
$ I $
is in terms of the ``bare'' parameters
$ \qh , A $
and
$ \Bh $.
We need an expression for
$ I $
in terms of the physical parameters
$  q $
(the magnetic charge ) and
$  B $
(the magnetic field of the Melvin metric that the Ernst metric is asymptotic
to).
We start by evaluating
$  B $.
Define the scalar
$ J $
to be the value of
$ F^2 $
on the axis (that is at
$ x = {\zeta_3} $).
In the Melvin Universe
$ J = 2 {\Bh^2} $
so in the Ernst metric
$ J $
far from the black holes will approach
$ 2 {{ B}^2} $.

Calculating
$ J $
we find
\eqn\gseven{\eqalign{
J &= {\lim _{x \to {\zeta_3}}} \;
\left [ 2 {g^{zz}} \; {g^{xx}} \; {{\left ( {F_{zx}}
\right ) }^2} \; + \; 2 {g^{zz}} \; {g^{yy}} \; {{\left ( {F_{zy}}
\right ) }^2} \right ]\cr
&= 2 \; {A^4} \; {{\left ( {\zeta_3} \, - \, y \right ) }^4} \; {{\left [ {\lim
_{x \to {\zeta_3}}} \; {{\partial E} \over {\partial x}} \right ] }^2}}
}
It then follows using the formula for
$ E $, \Edef, (and some straightforward but tedious algebra) that
\eqn\geight{
J = 2 {{\left [ \qh \; {{\left ( 1 + {1 \over 2} \qh \Bh {\zeta_3}
\right ) }^{ - 2}} \; {A^2} \; {{\left ( {\zeta_3} \, - \, y \right ) }^2}  +
  \Bh
{{\left ( 1  +  {1 \over 2}  \qh \Bh {\zeta_3} \right ) }^{ - 3}}
{\sqrt { 1 \, - \, 4 \qh A}} \right ] }^2} \ .}
We then find
\eqn\gnine{\eqalign{
{ B} &= {\lim _{y \to {\zeta_3}}} \; {\sqrt { J / 2}}\cr
&= \Bh \;  {{\left ( 1 \; + \; {1 \over 2} \; \qh \Bh {\zeta_3}
\right ) }^{ - 3}} \; {\sqrt { 1 \, - \, 4 \qh A}} \; \; \; .}}

Next we need to find $q$ in terms of the ``bare'' parameters
$ A, \Bh $
and
$ \qh$.
We have
\eqn\gten{
{ q} = {1 \over {4 \pi}} \; {\int _{S^2}} \; { F} }
where the integral is over the horizon 2-sphere.  We then find
\eqn\geleven{\eqalign{
{ q} =& {1 \over {4 \pi}} \; {\int _{\zeta_3} ^{\zeta_4}} \; d x \; {\int
_{z_-}
^{z_+}} \; d z \; {F_{x z}}\cr
=& {1 \over {4 \pi}} \; {\int _{z_-} ^{z_+}} \; d z \;  \; {\int _{\zeta_3}
^{\zeta_4}} \; d x \; \left ( - \; {{\partial E} \over {\partial x}} \right
)\cr
=& {{\Delta z} \over {4 \pi}} \; \left [ E ( {\zeta_3} ) \; - \; E ( {\zeta_4}
)
\right ]\cr
=& {{\Delta z} \over {4 \pi}} \; {2 \over \Bh} \; \left [ {{\left ( 1 \; +
\; {1 \over 2} \qh \Bh {\zeta_3} \right ) }^{ - 1}} \; - \;
{{\left ( 1 \; + \; {1 \over 2} \qh \Bh {\zeta_4} \right ) }^{ - 1}} \right ]\
.}}
However one can show that smoothness of the horizon two-sphere metric requires
\eqn\gtwelve{
\Delta z = {{2 \pi } \over {\sqrt {1 \, - \, 4 \qh A }}} \; {{\left ( 1 \; + \;
{1 \over 2} \qh \Bh {\zeta_3} \right ) }^{ 4}} \; \; \; .
}
So we find
\eqn\gthirt{\eqalign{
{ q} &= {1 \over {\Bh {\sqrt {1 \, - \, 4 \qh A }}}} \; {{\left ( 1 \; + \;
{1 \over 2} \qh \Bh {\zeta_3} \right ) }^{ 3}} \; \left [ 1 \; - \; {{\left ( 1
\; + \; {1 \over 2} \qh \Bh {\zeta_3} \right ) } \over {\left ( 1 \; + \;
{1 \over 2} \qh \Bh {\zeta_4} \right ) }} \right ]\cr
&= {1 \over { B}} \; \left [ 1 \; - \; {{ 1 \; + \; {1 \over 2} \qh \Bh
{\zeta_3}  } \over { 1 \; + \; {1 \over 2} \qh \Bh {\zeta_4} }} \right ] \; \;
\; .}}

We now find the constraint on the three bare parameters.  Smoothness of the
horizon two-sphere metric requires
\eqn\gfourt{
\left | {{G ' ( {\zeta_4} )} \over {{\Lambda ^2} ( {\zeta_4} ) }} \right | =
\left | {{G ' ( {\zeta_3} )} \over {{\Lambda ^2} ( {\zeta_3} ) }} \right |  \;
\; \; .
}
However we have
\eqn\gfift{
{\zeta_{3,4}} = {1 \over {2 \qh A }} \; \left ( - 1 \; + \; {\sqrt { 1 \, \mp 4
\qh A }}
\right ) \; \; \; .
}
The constraint then becomes
\eqn\gsixt{
1 = \left ( {{1 \; + \; 4 \qh A } \over {1 \; - 4 \qh A }} \right ) \; {{\left
(
{{ 1 \; + \; {1 \over 2} \qh \Bh {\zeta_3}  } \over { 1 \; + \; {1 \over 2} \qh
\Bh {\zeta_4} }} \right ) }^8} \; \; \; .
}

We next find an expression for
$ \qh A $
in terms of the physical parameters.  First define the parameter
$ u $
by
\eqn\gsevent{
u \equiv 1 \; - \; { B} { q} \; \; \; .
}
Then the expression for
$  q $
becomes
\eqn\geighte{
u =  {{ 1 \; + \; {1 \over 2} \qh \Bh {\zeta_3}  } \over { 1 \; + \; {1 \over
2} \qh \Bh {\zeta_4} }} \; \; \; .}
The constraint then becomes
\eqn\gninet{
{u^8} = {{1 \; - \; 4 \qh A } \over {1 \; + 4 \qh A }} }
which yields
\eqn\gtwent{
4 \qh A = {{1 \, - \, {u^8}} \over {1 \, + \, {u^8}}} \; \; \; .}

We now find an expression for
$ \Bh / A $
in terms of the physical parameters.  We have
\eqn\gtfive{
1 \; + \; {1 \over 2} \qh \Bh {\zeta_3}  = u \left ( 1 \; + \; {1 \over 2} \qh
\Bh {\zeta_4}
\right ) \; \; \; .}
Substituting the expressions for
$ \zeta_3$, $\zeta_4$
we find
\eqn\gtsix{
1 \; + {\Bh \over {4 A}} \; \left ( - 1 \; + {\sqrt {1 \, - \, 4 \qh A }}
\right ) = u \; + \; u \; {\Bh \over {4 A}} \; \left ( - 1 \; + {\sqrt {1 \, +
\, 4 \qh A }} \right ) \; \; \; .}
Rearranging terms we have
\eqn\gtseven{
1 = {\Bh \over {4 A}} \; \left [ 1 \; + \; {{u {\sqrt {1 \, + \, 4 \qh A }}
\; - \; {\sqrt {1 \, - \, 4 \qh A }} } \over { 1 \, - \, u }} \right ] \; \; \;
.}
Now substituting the expression for
$ \qh A $
we find
\eqn\gteight{
1 = {\Bh \over {4 A}} \; \left [ 1 \; + \; {\sqrt {2 \over {1 \, + \, {u^8}}}}
\; u \left ( {u^2} \, + \, u \, + 1 \right ) \right ] \; \; \; .
}
Define the quantity
$ f $
by
\eqn\gtnine{
f \equiv 1 \; + \; {\sqrt {2 \over {1 \, + \, {u^8}}}}
\; u \left ( {u^2} \, + \, u \, + 1 \right ) \; \; \; .
}
Then we have
\eqn\gthirty{
{\Bh \over A} = {4 \over f} \; \; \; .
}

We now find an expression for
$ 1 / A $
in terms of the physical parameters.
First we have
\eqn\gthone{\eqalign{
1 \; + \; {1 \over 2} \; \qh \Bh {\zeta_3}  &= 1 \; + \; {\Bh \over {4 A}} \;
\left (
- 1 \; + \; {\sqrt {1 \, - \, 4 \qh A}} \right )\cr
&= 1 \; + {f^{- 1}} \left ( - 1 \; + \; {\sqrt {2 \over {1 \, + \, {u^8}}}} \;
{u^4} \right )\cr
&= {f^{ - 1}} \; \left ( f \; - \; 1 \; + \; {\sqrt
{2 \over {1 \, + \, {u^8}}}} \; {u^4} \right ) \cr
&= {f^{ - 1}} \; {\sqrt {2 \over {1 \, + \, {u^8}}}}
\; u \; {{1 \, - \, {u^4}}
\over {1 \, - \, u}} \; \; \; .}}
It then follows that
\eqn\gthtwo{\eqalign{
{1 \over A} &= {1 \over { B}} \; {{ B} \over \Bh} \;  {\Bh \over A}\cr
&= {1 \over { B}} \; {{\left ( 1 \; + \; {1 \over 2} \; \qh \Bh {\zeta_3}
\right )
}^{- 3}} \; {\sqrt {1 \, - \, 4 \qh A }} \; {4 \over f}\cr
&= {1 \over { B}}  \; 2 {f^2} \; u \left ( 1 \, + \, {u^8} \right ) \;
{{\left ( {{1 \, - \, u} \over {1 \, - \, {u^4} }} \right ) }^3} \; \; \; .}}

We are now ready to evaluate the quantity
$ I $.
\eqn\gththree{\eqalign{
I &= 2 \pi \, \left ( {1 \over A} \; - \; {\Bh \over {2 {A^2}}} \; - \; {{\qh
{\Bh^2}}
\over {4 {A^2}}} \right ) \cr
&= {{2 \pi} \over A} \; \left ( 1 \; - \; {\Bh \over {2 A}} \; - \; {1 \over 4}
\;
\qh A \; {{\left [ {\Bh \over A} \right ] }^2} \right ) \cr
&= {{2 \pi} \over { B}} \; 2 {f^2} \; u \left ( 1 \, + \, {u^8} \right ) \;
{{\left ( {{1 \, - \, u} \over {1 \, - \, {u^4} }} \right ) }^3} \; \left (
1 \; - \; {2 \over f} \; - \; {{1 \, - \, {u^8}} \over {1 \, + \, {u^8}}}
\; {1 \over {f^2}} \right ) \cr
&= {{4 \pi} \over { B}} \; u \left ( 1 \, + \, {u^8} \right ) \; {{\left (
{{1 \, - \, u} \over {1 \, - \, {u^4} }} \right ) }^3} \;
\left ( {{[f \, - \, 1]}^2} \; - \; 1 \; - \; {{1 \, - \, {u^8}}
\over {1 \, + \, {u^8}}} \right )\cr
&= {{8 \pi} \over { B}} \; {{u \left ( {u^3} \, + \, {u^2} \, + \, u \, -
\, 1 \right ) } \over {{\left ( {u^3} \, + \, {u^2} \,
 + \, u \, + \, 1 \right ) }^2}} \; \; \; .
}}

Now we are (finally) ready to evaluate the action.
\eqn\gthfour{\eqalign{
S &= {\int _0 ^{ q}} \; I \; d { q}\cr
&= {1 \over { B} } \; {\int _{1 - { B} { q}} ^1} \; I \; d u
\; \; \; .}}
The integral is
\eqn\gthfive{\eqalign{
S &= {{4 \pi} \over {{ B}^2}} \; {{\left [ {{2 {u^2} \, + \, u \, + \, 1}
\over {{u^3} \, + \, {u^2} \, + \, u \, + \, 1 }} \right ] }_1 ^{1 - { B}
{ q}}}\cr
&= {{4 \pi} \over {{ B}^2}} \; {{\left [ 1 \; + {{{u^2} \, - \, {u^3}} \over
{{u^3} \, + \, {u^2} \, + \, u \, + \, 1 }} \right ] }_1 ^{1 - { B}
{ q}}}\cr
&= {{4 \pi} \over {{ B}^2}} \; {{\left [ 1 \; + {{{u^2} \; {{( 1 \, - \,
u)}^2}} \over {1 \, - \, {u^4}}} \right ] }_1 ^{1 - { B}
{ q}}} \cr
&= 4 \pi {{ q}^2} \; {{{\left ( 1 \, - \, { B} { q} \right ) }^2} \over {1 \; -
\; {{\left ( 1 \, - \, { B} { q} \right ) }^4}}}
\ ,}}
as quoted in \instact.

\bigskip\bigskip\centerline{{\bf Acknowledgements}}\nobreak
We thank L. Susskind for instigating these investigations by
asking about the role of black hole entropy in
pair production.  We are grateful to M. O'Loughlin for supplying the
figures.
This work was initiated at the Aspen Center for Physics,
whose hospitality we acknowledge, and was supported in part by
DOE grant DOE-91ER40618,  NSF grant PHY89-04035, and NSF PYI grant
PHY-9157463 to SBG.

\listrefs

\end